\documentclass[aps,preprint]{revtex4}%
\usepackage{amsfonts}
\usepackage{amsmath}
\usepackage{amssymb}
\usepackage{graphicx}%
\providecommand{\U}[1]{\protect\rule{.1in}{.1in}}
\newtheorem{theorem}{Theorem}
\newtheorem{acknowledgement}[theorem]{Acknowledgement}

\begin{document}
\preprint{ }
\title{Different Electronic Charges in Two-Component Superconducter by Coherente State}
\author{Xuguang Shi}
\affiliation{College of Science, Beijing Forestry University, Beijing 100083, P.R.China}
\keywords{}
\pacs{}

\begin{abstract}
\ Recently, the different electronic charges, which are related to the
different coupling constants with magnetic field, in the two-component
superconductor have been studied in frame of Ginzburg-Landau theory. In order
to study the electronic charges in detail we suggest the wave function in the
two-component superconductor to be coherent state. We find the different
electronic charges exist not only in the coherent state but the incoherent
state. But the ratio of the different charges in the coherent state is
different from ratio in the incoherence. The expressions of the coupling
constants are given directly based on the coherence effects. We also discuss
the winding number in such system.

\textbf{Keywords:} two-component superconductor; Ginzburg-Landau theory;
winding number; cohernet state

\textbf{PACS: }02.40.Pc; 67.85.Fg; 74.25.Ha

\end{abstract}
\startpage{1}
\maketitle

\section*{Introduction}

Two-component superconductor \cite{two component,two component1,two
component2,twocom} has inspired great interests since high temperature
superconductor $MgB_{2}$ was observed in the experiments \cite{MgB,MgB1,MgB2}.
The reason is that the two-component superconductor exhibits some novel
properties. For example, there are composite vortices structure
\cite{compositV,compositV1,compositV2}; The two-component superconductor is
very different from the type I and type II superconductor. In Ref
\cite{Typ1.5,Typ1.51}, the authors call it type 1.5 superconductor. Recently,
the different electronic charges carried by the condensates are studied by
using of Ginzburg-Landau theory \cite{different charges}. They assume that the
coupling constants in the two-component superconductor are different. The
coupling constants come from the interaction between the superpositions of the
condensates and magnetic field. The electronic charges are expressed by the
product of the coupling constants and electronic charges carried by Cooper
pairs. This is the reason leading to the different electronic charges
\cite{different charges}. The ratio of the two coupling constants is just the
ratio of the two different electronic charges, which is rational number. In
Ref \cite{different charges}, they find that the different condensates carry
the different windings and different fractional vortices by using of the flux
quantization and finiteness of the energy per unit length.

The different electric charges can exist in the multicomponent systems, such
as two-component superconductor and liquid metallic deuterium. We also suggest
the different electric charges exist in the Josephson effect \cite{JE}. The
reason leading to the different electric charges is the mixtures of
condensates. In general, several superpositions are need to describe the
condensates in such systems. In order to study the mixtures, it is nature to
assume the superpositions to be in coherence, or linear combination. In this
letter, we suggest the superpositions of the two condensates in the
two-component superconductor are coherent. We devote to study how the
coherence effects give rise to the different coupling constants. We also try
to find how the coherent state and incoherent state affect the two electronic
charges. The ratio of the different charges are studied in the coherent state
and incoherent state and we find the ratios are different. Based on $\phi
-$mapping topological theory \cite{topo,topo1,topo2}, we discuss which
condition leads to integer vorticity and which condition leads to fractional
vorticity. The contents are arranged as follows: in section 2, we suggest the
superpositions describing the two condensates are coherent. The different
electronic charges are studied by using of this coherent state. The
expressions of the coupling constants in the coherent state and incoherent
state are deduced directly. In section 3, in the frame of $\phi-$mapping
topological theory, we find the velocity of the coherent state can carry
magnetic flux. The vorticities of the velocity in the coherent state and
incoherent state are expressed by winding number. Then the fractional
vorticity and magnetic flux are obtained. In section 4, a conclusion is given.

\section{Different electronic charges in the two-component superconductor by
coherent state}

The Ginzburg-Landau free energy, which is used to describe the two-component
condensate, is expressed as%
\begin{equation}
F=\sum_{\alpha}\frac{\hbar^{2}}{2m}\left[  \left(  \nabla+i\frac{e}{\hbar
c}\mathbf{A}\right)  \psi_{\alpha}\right]  ^{2}+V\left(  \psi\right)
+\frac{1}{8\pi}\mathbf{B}^{2},\text{ \ }\alpha=1,2\label{0}%
\end{equation}
where $e=2e^{\ast}$ is the electronic charge of the Cooper pair. Here,
$e^{\ast}$ is the electronic charge. $\mathbf{A}$ is $U(1)$ gauge potential
and $\mathbf{B}$ is magnetic field%
\begin{equation}
\mathbf{B}=\nabla\times\mathbf{A}.
\end{equation}
The potential $V$ is%
\begin{equation}
V=\sum_{\alpha}\left[  a_{\alpha}\left\vert \psi_{\alpha}\right\vert
^{2}+\frac{1}{2}b_{\alpha}\left\vert \psi_{\alpha}\right\vert ^{4}\right]
.\label{Vp}%
\end{equation}
where $\alpha=1,2.$ Let us assume the superpositions $\psi_{1}$ and $\psi_{2}$
describe two condensates respectively. Then we suggest the total wave function
in the two-component superconductor should be coherent state and write it as%
\begin{equation}
\Psi=C_{1}\psi_{1}+C_{2}\psi_{2}.\label{1}%
\end{equation}
The superposition of every condensate is the anzat $\psi_{\alpha}=\sqrt
{\rho_{\alpha}}e^{i\theta_{\alpha}}.$ Then the densities of the superpositions
respectively are%
\begin{equation}
\rho_{\alpha}=\psi_{\alpha}\psi_{\alpha}^{\ast}\text{ \ \ }\alpha
=1,2.\label{2}%
\end{equation}
By considering the condition of normalization, we have%
\begin{equation}%
{\displaystyle\sum\limits_{\alpha=1}^{2}}
C_{\alpha}C_{\alpha}^{\ast}=1.
\end{equation}
The density of the coherent state is%
\begin{equation}
\rho=\Psi\Psi^{\ast}=C_{1}C_{1}^{\ast}\psi_{1}\psi_{1}^{\ast}+C_{2}C_{2}%
^{\ast}\psi_{2}\psi_{2}^{\ast}+C_{1}C_{2}^{\ast}\psi_{1}\psi_{2}^{\ast}%
+C_{2}C_{1}^{\ast}\psi_{2}\psi_{1}^{\ast}.\text{ }%
\end{equation}
This density also can be written as%
\begin{equation}
\rho=C_{1}C_{1}^{\ast}\psi_{1}\psi_{1}^{\ast}+C_{2}C_{2}^{\ast}\psi_{2}%
\psi_{2}^{\ast}+2\operatorname{Re}\left[  C_{1}C_{2}^{\ast}\psi_{1}\psi
_{2}^{\ast}\right]  .\label{den}%
\end{equation}
When the wave function $\Psi$ is in coherent state, the term in (\ref{den})
$\operatorname{Re}\left[  C_{1}C_{2}^{\ast}\psi_{1}\psi_{2}^{\ast}\right]
\neq0$, which exhibits coherent effect. When the wave function $\Psi$ is in
incoherent state, the term $\operatorname{Re}\left[  C_{1}C_{2}^{\ast}\psi
_{1}\psi_{2}^{\ast}\right]  =0$, the coherent effect of the system disappears.
Support the Schr\"{o}dinger equation of a charged particle without spin in a
electromagnetic field is%
\begin{equation}
i\hbar\frac{\partial\psi}{\partial t}=H\psi,
\end{equation}
where $H=-\frac{\hbar^{2}}{2m}\left(  \nabla+i\frac{e}{\hbar c}\mathbf{A}%
\right)  ^{2}+V$. Let $\psi_{1}$ and $\psi_{2}$ be the solutions of the
Schr\"{o}dinger equation, then the coherent state $\Psi$ is also the solution
of the Schr\"{o}dinger equation. It should be noted that the potential $V$ is
real number and it has no contribution to probability current. Then the
probability current of the coherent state satisfies%
\begin{equation}
\delta H=-\frac{1}{c}\int\mathbf{J\cdot}\delta\mathbf{A}dV.
\end{equation}
Then the probability current is
\begin{equation}
\mathbf{J}=\frac{i\hbar}{2m}\left(  \Psi^{\ast}\left(  \mathbf{\nabla}%
+i\frac{e}{\hbar c}\mathbf{A}\right)  \Psi-\Psi\left(  \mathbf{\nabla}%
+i\frac{e}{\hbar c}\mathbf{A}\right)  \Psi^{\ast}\right)  ,
\end{equation}
which is just the superconducting current. Put (\ref{1}) into this formula, we
have%
\begin{align}
\mathbf{J} &  =\frac{i\hbar}{2m}\sum_{\alpha=1}^{2}C_{\alpha}C_{\alpha}^{\ast
}\left(  \psi_{\alpha}^{\ast}\mathbf{\nabla}\psi_{\alpha}-\psi_{\alpha
}\mathbf{\nabla}\psi_{\alpha}^{\ast}\right)  \nonumber\\
&  +\frac{i\hbar}{2m}C_{1}C_{2}^{\ast}\left(  \psi_{2}^{\ast}\mathbf{\nabla
}\psi_{1}-\psi_{1}\mathbf{\nabla}\psi_{2}^{\ast}\right)  +\frac{i\hbar}%
{2m}C_{1}C_{2}^{\ast}\left(  \psi_{1}^{\ast}\mathbf{\nabla}\psi_{2}-\psi
_{2}\mathbf{\nabla}\psi_{1}^{\ast}\right)  -\frac{e}{mc}\mathbf{A}\Psi
\Psi^{\ast}.
\end{align}
Substitute the electronic charge into this formula, the current will become to
the electronic current. We write it as%
\begin{align}
\mathbf{J}_{e} &  =\frac{i\hbar e}{2m}\sum_{\alpha=1}^{2}C_{\alpha}C_{\alpha
}^{\ast}\left(  \psi_{\alpha}^{\ast}\mathbf{\nabla}\psi_{\alpha}-\psi_{\alpha
}\mathbf{\nabla}\psi_{\alpha}^{\ast}\right)  \nonumber\\
&  +\frac{i\hbar e}{2m}C_{1}C_{2}^{\ast}\left(  \psi_{2}^{\ast}\mathbf{\nabla
}\psi_{1}-\psi_{1}\mathbf{\nabla}\psi_{2}^{\ast}\right)  +\frac{i\hbar e}%
{2m}C_{1}C_{2}^{\ast}\left(  \psi_{1}^{\ast}\mathbf{\nabla}\psi_{2}-\psi
_{2}\mathbf{\nabla}\psi_{1}^{\ast}\right)  -\frac{e^{2}}{mc}\mathbf{A}\Psi
\Psi^{\ast}.
\end{align}
The second and third terms show the coherence effects of the two
superpositions and we denote these terms as $\mathbf{J}_{e2}$. The other terms
are denoted as $\mathbf{J}_{e1}.$
\begin{align}
\mathbf{J}_{e1} &  =\frac{i\hbar e}{2m}\sum_{\alpha=1}^{2}C_{\alpha}C_{\alpha
}^{\ast}\left(  \psi_{\alpha}^{\ast}\mathbf{\nabla}\psi_{\alpha}-\psi_{\alpha
}\mathbf{\nabla}\psi_{\alpha}^{\ast}\right)  -\frac{e^{2}}{mc}\mathbf{A}%
\Psi\Psi^{\ast},\label{je1}\\
\mathbf{J}_{e2} &  =\frac{i\hbar e}{2m}C_{1}C_{2}^{\ast}\left(  \psi_{2}%
^{\ast}\mathbf{\nabla}\psi_{1}-\psi_{1}\mathbf{\nabla}\psi_{2}^{\ast}\right)
+\frac{i\hbar e}{2m}C_{1}C_{2}^{\ast}\left(  \psi_{1}^{\ast}\mathbf{\nabla
}\psi_{2}-\psi_{2}\mathbf{\nabla}\psi_{1}^{\ast}\right)  .\label{je2}%
\end{align}
Let us define new complex variable as%
\begin{equation}
\Lambda=C_{1}C_{2}^{\ast}\psi_{1}\psi_{2}^{\ast},\label{P1}%
\end{equation}
then the complex conjugation of this variable is%
\begin{equation}
\Lambda^{\ast}=C_{1}^{\ast}C_{2}\psi_{1}^{\ast}\psi_{2}.\label{P2}%
\end{equation}
These complex numbers also can be rewritten as%
\begin{align}
\Lambda &  =\Lambda_{1}+i\Lambda_{2},\nonumber\\
\Lambda^{\ast} &  =\Lambda_{1}-i\Lambda_{2},\label{P3}%
\end{align}
where $\Lambda_{1}$ and $\Lambda_{2}$ are real numbers. Put the formulas
(\ref{P1},\ref{P2},\ref{P3}) into the density (\ref{den}), we get the density
of the wave function%
\begin{equation}
\rho=\left\vert C_{1}\right\vert ^{2}\rho_{1}+\left\vert C_{2}\right\vert
^{2}\rho_{2}+2\Lambda_{1}.
\end{equation}
These parameters provide us one way to determine the coherence effects of the
system. When $\Lambda=0$, the wave function in the two-component
superconductor is the incoherent state. When $\Lambda\neq0,$ the wave function
is the coherent state. From (\ref{je2}), the electronic current $\mathbf{J}%
_{e2}$ is deduced as%
\begin{equation}
\mathbf{J}_{e2}=\frac{i\hbar e}{2m}\Lambda_{1}\left[  \frac{\left(  \psi
_{1}^{\ast}\mathbf{\nabla}\psi_{1}-\psi_{1}\mathbf{\nabla}\psi_{1}^{\ast
}\right)  }{\psi_{1}\psi_{1}^{\ast}}+\frac{\left(  \psi_{2}^{\ast
}\mathbf{\nabla}\psi_{2}-\psi_{2}\mathbf{\nabla}\psi_{2}^{\ast}\right)  }%
{\psi_{2}\psi_{2}^{\ast}}\right]  -\frac{\hbar e\Lambda_{2}}{2m}%
\mathbf{\nabla}\ln\left(  \frac{\psi_{1}\psi_{1}^{\ast}}{\psi_{2}\psi
_{2}^{\ast}}\right)  .
\end{equation}
The electronic current $\mathbf{J}_{e1}$ can be given as%
\begin{equation}
\mathbf{J}_{e1}=\frac{i\hbar e}{2m}\sum_{\alpha}\left\vert C_{\alpha
}\right\vert ^{2}\rho_{\alpha}\left(  \frac{\psi_{\alpha}^{\ast}%
\mathbf{\nabla}\psi_{\alpha}-\psi_{\alpha}\mathbf{\nabla}\psi_{\alpha}^{\ast}%
}{\psi_{\alpha}\psi_{\alpha}^{\ast}}\right)  -\frac{e^{2}}{mc}\mathbf{A}%
\Psi\Psi^{\ast}.
\end{equation}
Finally, the total electronic current is%
\begin{align}
\mathbf{J}_{e} &  =\frac{i\hbar e}{2m}\left(  \left\vert C_{1}\right\vert
^{2}\rho_{1}+\Lambda_{1}\right)  \frac{\left(  \psi_{1}^{\ast}\mathbf{\nabla
}\psi_{1}-\psi_{1}\mathbf{\nabla}\psi_{1}^{\ast}\right)  }{\psi_{1}\psi
_{1}^{\ast}}+\frac{i\hbar e}{2m}\left(  \left\vert C_{2}\right\vert ^{2}%
\rho_{2}+\Lambda_{1}\right)  \frac{\left(  \psi_{2}^{\ast}\mathbf{\nabla}%
\psi_{2}-\psi_{2}\mathbf{\nabla}\psi_{2}^{\ast}\right)  }{\psi_{2}\psi
_{2}^{\ast}}\nonumber\\
&  -\frac{\hbar e\Lambda_{2}}{2m}\mathbf{\nabla}\ln\left(  \frac{\rho_{1}%
}{\rho_{2}}\right)  -\frac{e^{2}}{mc}\mathbf{A}\Psi\Psi^{\ast}.\label{rr}%
\end{align}
It is should be noted that the term $\mathbf{\nabla}\ln\left(  \frac{\rho_{1}%
}{\rho_{2}}\right)  $ is vector and $\ln\left(  \frac{\rho_{1}}{\rho_{2}%
}\right)  $ can be denoted as $\theta$. By using of these definitions, the
vector is%
\begin{equation}
\mathbf{\nabla}\theta\left(  x\right)  =\mathbf{\nabla}\ln\left(  \frac
{\rho_{1}}{\rho_{2}}\right)  .
\end{equation}
When the wave function in the two-component superconductor is the incoherent
state, $\Lambda_{1}=\Lambda_{2}=0,$ the term $\mathbf{\nabla}\theta\left(
x\right)  $ disappears. The electronic current is same as the current obtained
by $\frac{\delta F}{\delta A}$. When the wave function is the coherent state,
$\Lambda_{1}\neq0,\Lambda_{2}\neq0,$ the term $\mathbf{\nabla}\theta\left(
x\right)  $ is analyzed as follows: if $\frac{\rho_{1}}{\rho_{2}}$ is a
function of coordinates and the function $\theta\left(  x\right)  $ satisfies
Clairaut's theorem, the second mixed derivatives of $\theta\left(  x\right)  $
is symmetry%
\begin{equation}
\partial_{i}\partial_{j}\theta\left(  x\right)  =\partial_{j}\partial
_{i}\theta\left(  x\right)  .\label{F}%
\end{equation}
This shows%
\begin{equation}
\mathbf{\nabla}\times\mathbf{\nabla}\theta\left(  x\right)  =0.\label{C}%
\end{equation}
Then the function $\theta\left(  x\right)  $ can be seen as the phase of the
$U(1)$ gauge magnetic potential, the $U(1)$ gauge potential is translated as
\begin{equation}
\mathbf{A}^{^{\prime}}\rightarrow\mathbf{A}+\mathbf{\nabla}\theta.
\end{equation}
Because of Eq.(\ref{C}), the magnetic field under this translation is
$\mathbf{B}^{^{\prime}}=\nabla\times\mathbf{A}^{^{\prime}}=\mathbf{B}$. The
phase has no contribution to the magnetic field. This result shows that the
$U\left(  1\right)  $-like phase $\theta$ does not lead to observed
phenomenon. It is natural to define the currents as $\mathbf{j}_{1}$ and
$\mathbf{j}_{2}$ respectively%
\begin{align}
\mathbf{j}_{1} &  =\rho_{1}\mathbf{V}_{1},\nonumber\\
\mathbf{j}_{2} &  =\rho_{2}\mathbf{V}_{2},\label{j}%
\end{align}
where%
\begin{align}
\mathbf{V}_{1} &  =\frac{\left(  \psi_{1}^{\ast}\mathbf{\nabla}\psi_{1}%
-\psi_{1}\mathbf{\nabla}\psi_{1}^{\ast}\right)  }{\psi_{1}\psi_{1}^{\ast}%
},\nonumber\\
\mathbf{V}_{2} &  =\frac{\left(  \psi_{2}^{\ast}\mathbf{\nabla}\psi_{2}%
-\psi_{2}\mathbf{\nabla}\psi_{2}^{\ast}\right)  }{\psi_{2}\psi_{2}^{\ast}}.
\end{align}
Then the total electronic current is%
\begin{equation}
\mathbf{J}_{e}=\frac{i\hbar e}{2m}\left(  \left\vert C_{1}\right\vert ^{2}%
\rho_{1}+\Lambda_{1}\right)  \mathbf{V}_{1}+\frac{i\hbar e}{2m}\left(
\left\vert C_{2}\right\vert ^{2}\rho_{2}+\Lambda_{1}\right)  \mathbf{V}%
_{2}-\frac{e^{2}}{mc}\mathbf{A}\Psi\Psi^{\ast}-\frac{\hbar e\Lambda_{2}}%
{2m}\mathbf{\nabla}\theta\left(  x\right)  .
\end{equation}
In order to study the electronic charges in the two-component superconductor,
we write the electronic current as%
\begin{equation}
\mathbf{J}_{e}=\frac{i\hbar e}{2m}\left(  \left\vert C_{1}\right\vert
^{2}+\frac{\Lambda_{1}}{\rho_{1}}\right)  \mathbf{j}_{1}+\frac{i\hbar e}%
{2m}\left(  \left\vert C_{2}\right\vert ^{2}+\frac{\Lambda_{1}}{\rho_{2}%
}\right)  \mathbf{j}_{2}-\frac{e^{2}}{mc}\mathbf{A}\Psi\Psi^{\ast}-\frac{\hbar
e\Lambda_{2}}{2m}\mathbf{\nabla}\theta\left(  x\right)  ,
\end{equation}
here, we use the relations (\ref{j}). When the wave function is the coherent
state, $\Lambda_{1}\neq0$, then the electronic charges in the superconductor
are%
\begin{align}
e_{1} &  =\left(  \left\vert C_{1}\right\vert ^{2}+\frac{\Lambda_{1}}{\rho
_{1}}\right)  e,\nonumber\\
e_{2} &  =\left(  \left\vert C_{2}\right\vert ^{2}+\frac{\Lambda_{1}}{\rho
_{2}}\right)  e.
\end{align}
The ratio of the two charges is%
\begin{equation}
\frac{e_{1}}{e_{2}}=\frac{\left(  \left\vert C_{1}\right\vert ^{2}%
+\frac{\Lambda_{1}}{\rho_{1}}\right)  }{\left(  \left\vert C_{2}\right\vert
^{2}+\frac{\Lambda_{1}}{\rho_{2}}\right)  }.\label{R1}%
\end{equation}
Where we find the coupling constants $g_{1}$ and $g_{2}$ in Ref.
\cite{different charges} must be%
\begin{align}
g_{1} &  =\left\vert C_{1}\right\vert ^{2}+\frac{\Lambda_{1}}{\rho_{1}%
},\nonumber\\
g_{2} &  =\left\vert C_{2}\right\vert ^{2}+\frac{\Lambda_{1}}{\rho_{2}}.
\end{align}
When the wave function is the incoherent state, $\Lambda_{1}=\Lambda_{2}=0$,
then the electronic current is%
\[
\mathbf{J}_{e}=\frac{i\hbar\left\vert C_{1}\right\vert ^{2}e}{2m}%
\mathbf{j}_{1}+\frac{i\hbar\left\vert C_{2}\right\vert ^{2}e}{2m}%
\mathbf{j}_{2}-\frac{e^{2}}{mc}\mathbf{A}\Psi\Psi^{\ast}.
\]
In this case the electronic charges are $e_{1}=\left\vert C_{1}\right\vert
^{2}e$ and $e_{2}=\left\vert C_{2}\right\vert ^{2}e$. Therefore, we find
$g_{1}$ and $g_{2}$ must be%
\begin{align}
g_{1} &  =\left\vert C_{1}\right\vert ^{2}\nonumber\\
g_{2} &  =\left\vert C_{2}\right\vert ^{2}.
\end{align}
The ratio of the two charges is%
\begin{equation}
\frac{e_{1}}{e_{2}}=\frac{\left\vert C_{1}\right\vert ^{2}}{\left\vert
C_{2}\right\vert ^{2}}.\label{R2}%
\end{equation}
Considering $\mathbf{J}_{e}=\rho\mathbf{V},$ the velocity of the wave function
is%
\begin{equation}
\mathbf{V}=\frac{i\hbar e}{2m}\frac{\left(  \left\vert C_{1}\right\vert
^{2}\rho_{1}+\Lambda_{1}\right)  }{\rho}\mathbf{V}_{1}+\frac{i\hbar e}%
{2m}\frac{\left(  \left\vert C_{2}\right\vert ^{2}\rho_{2}+\Lambda_{1}\right)
}{\rho}\mathbf{V}_{2}-\frac{e^{2}}{mc}\mathbf{A}-\frac{\hbar e\Lambda_{2}%
}{2m\rho}\mathbf{\nabla}\theta\left(  x\right)  .\label{veloc}%
\end{equation}
\ All these show that the different electronic charges in the two-component
superconductor exist not only in the coherent state but the incoherent state.
But from (\ref{R1}) and (\ref{R2}), the ratio of the difference electronic
charges in the coherence is different from the the ratio in the incoherence.
The same analyze can be done for the coupling constants. Although the
interaction between two condensates are mediated by the $U(1)$ magnetic field,
the reason leading to different electronic charges, or coupling constants, is
that there are coherence effects in the two-component superpositions. It is
interesting that when the wave function is in the coherence, there is a
$U(1)$-like phase adding to the gauge potential $\mathbf{A}$. This phase has
no contribution to magnetic field. When the wave function is the incoherent
state, the phase disappears and has no contribution to magnetic field too. So
the coherence effects do not affect the magnetic field.

\section{Winding number in the two-component superconductor}

Topology has play an important role in mathematics and physics. Topological
defect, or phase singularity, appears in many physics system. Based on the
$\phi-$mapping topological theory, the direct and explicit expression of the
isolated singular topological structures in the continuous media have been
mathematically deduced, such as monopoles \cite{topo}, superconductor
\cite{compositV2}, Chern-Simons vortex \cite{duan}. In $\phi-$mapping
topological theory, the wave function $\phi\left(  x^{i}\right)  $ $(i=1,2,3)$
is 2-dimensional vector field can be looked upon as a smooth map
$\phi:X\rightarrow R^{2}$. $X$ is a 3-dimensional Riemannian manifold and
$R^{2}$ is a 2-dimensional Euclidean space. When $\phi=0$ at point $p$, the
topological properties of this map are nontrivial. The zero point $p$ is
called a singular point of $\phi$. By using of $\phi-$mapping topological
theory, a topological number, called winding number, is obtained. It can be
used to describe the topological properties of singular point. The
superpositions of the two condensates are expressed as%
\begin{align}
\psi_{1}  &  =\psi_{1}^{1}+i\psi_{1}^{2},\nonumber\\
\psi_{2}  &  =\psi_{2}^{1}+i\psi_{2}^{2}.
\end{align}
By using of $\phi-$mapping topological theory, the unit vectors $n^{a}$ must
be introduced.%
\begin{align}
n_{1}^{a}  &  =\frac{\psi_{1}^{a}}{\left\Vert \psi_{1}\right\Vert
},\nonumber\\
n_{2}^{a}  &  =\frac{\psi_{2}^{a}}{\left\Vert \psi_{2}\right\Vert }\text{
\ \ }a=1,2.
\end{align}
It is found that the unit vectors exhibit some singularities at where the wave
function is zero. At the zero point, the phase of the wave function is
uncertain. So the nontrivial topological properties can be deduced from the
unit vectors directly \cite{topo,topo2}. The unit vectors should satisfy
$n_{1(2)}^{a}n_{1(2)}^{a}=1$. The velocities of the two superpositions are
calculated from these unit vectors as%
\begin{align}
V_{1i}  &  =\frac{\hbar}{m}\epsilon_{ab}n_{1}^{a}\partial_{i}n_{1}%
^{b},\nonumber\\
V_{2i}  &  =\frac{\hbar}{m}\epsilon_{ab}n_{2}^{a}\partial_{i}n_{2}^{b}.
\end{align}
The curl of the velocity is%
\begin{align}
\mathbf{\nabla}\times\mathbf{V}  &  =\frac{i\hbar e}{2m}\frac{\left(
\left\vert C_{1}\right\vert ^{2}\rho_{1}+\Lambda_{1}\right)  }{\rho
}\mathbf{\nabla}\times\mathbf{V}_{1}+\frac{i\hbar e}{2m}\frac{\left(
\left\vert C_{2}\right\vert ^{2}\rho_{2}+\Lambda_{1}\right)  }{\rho
}\mathbf{\nabla}\times\mathbf{V}_{2}\nonumber\\
&  -\frac{e^{2}}{mc}\mathbf{\nabla}\times\mathbf{A}-\frac{\hbar e\Lambda_{2}%
}{2m\rho}\mathbf{\nabla}\times\mathbf{\nabla}\theta\left(  x\right)  .
\end{align}
By considering (\ref{C}), the curl of the velocity is%
\begin{equation}
\left(  \mathbf{\nabla}\times\mathbf{V}\right)  _{i}=\frac{\hbar e}{m}%
\frac{\left(  \left\vert C_{1}\right\vert ^{2}\rho_{1}+\Lambda_{1}\right)
}{\rho}\epsilon^{ijk}\epsilon_{ab}\partial_{j}n_{1}^{a}\partial_{k}n_{1}%
^{b}+\frac{\hbar e}{m}\frac{\left(  \left\vert C_{2}\right\vert ^{2}\rho
_{2}+\Lambda_{1}\right)  }{\rho}\epsilon^{ijk}\epsilon_{ab}\partial_{j}%
n_{2}^{a}\partial_{k}n_{2}^{b}-\frac{e^{2}}{mc}\left(  \mathbf{\nabla}%
\times\mathbf{A}\right)  _{i}.
\end{equation}
We find that there is magnetic flux carried by the velocity through rewriting
the curl of the velocity as%
\begin{equation}
\left(  \mathbf{\nabla}\times\mathbf{V}\right)  _{i}=\frac{e^{2}}{mc}\left[
\frac{\hbar c}{e}\frac{\left(  \left\vert C_{1}\right\vert ^{2}\rho
_{1}+\Lambda_{1}\right)  }{\rho}\epsilon^{ijk}\epsilon_{ab}\partial_{j}%
n_{1}^{a}\partial_{k}n_{1}^{b}+\frac{\hbar c}{e}\frac{\left(  \left\vert
C_{2}\right\vert ^{2}\rho_{2}+\Lambda_{1}\right)  }{\rho}\epsilon
^{ijk}\epsilon_{ab}\partial_{j}n_{2}^{a}\partial_{k}n_{2}^{b}-\left(
\mathbf{\nabla}\times\mathbf{A}\right)  _{i}\right]  .
\end{equation}
The vorticity of the velocity (\ref{veloc}) will be%
\begin{align}
\int\left(  \mathbf{\nabla}\times\mathbf{V}\right)  \cdot d\mathbf{S}  &
=\frac{e^{2}}{mc}\left[  \frac{\hbar c}{e}\frac{\left(  \left\vert
C_{1}\right\vert ^{2}\rho_{1}+\Lambda_{1}\right)  }{\rho}\int\epsilon
^{ijk}\epsilon_{ab}\partial_{j}n_{1}^{a}\partial_{k}n_{1}^{b}\mathbf{e}%
_{i}\cdot d\mathbf{S}\right] \nonumber\\
&  +\frac{e^{2}}{mc}\left[  \frac{\hbar c}{e}\frac{\left(  \left\vert
C_{2}\right\vert ^{2}\rho_{2}+\Lambda_{1}\right)  }{\rho}\int\epsilon
^{ijk}\epsilon_{ab}\partial_{j}n_{2}^{a}\partial_{k}n_{2}^{b}\mathbf{e}%
_{i}\cdot d\mathbf{S}\right]  +\frac{e^{2}}{mc}\int\left(  \mathbf{\nabla
}\times\mathbf{A}\right)  \cdot d\mathbf{S}.
\end{align}
We note that the integral $\int\left(  \mathbf{\nabla}\times\mathbf{A}\right)
\cdot d\mathbf{S=}\int$ $\mathbf{B\cdot}d\mathbf{S}$ in the last term is
magnetic flux. The magnetic flux quanta is $\Phi_{0}=\frac{hc}{e}$, then the
magnetic flux can be expressed as%
\begin{equation}
\Phi=W^{^{\prime\prime}}\Phi_{0}=W^{^{\prime\prime}}\frac{hc}{e}.
\end{equation}
According $\phi-$mapping topological theory, the first term and second term
will be
\begin{align}
\int\left(  \mathbf{\nabla}\times\mathbf{V}_{1}\right)  \cdot d\mathbf{S}  &
=\int\epsilon^{ijk}\epsilon_{ab}\partial_{j}n_{1}^{a}\partial_{k}n_{1}%
^{b}\mathbf{e}_{i}\cdot d\mathbf{S}=2\pi W_{1},\nonumber\\
\int\left(  \mathbf{\nabla}\times\mathbf{V}_{2}\right)  \cdot d\mathbf{S}  &
=\int\epsilon^{ijk}\epsilon_{ab}\partial_{j}n_{2}^{a}\partial_{k}n_{2}%
^{b}\mathbf{e}_{i}\cdot d\mathbf{S}=2\pi W_{2}, \label{VC}%
\end{align}
where $W$ is called the winding number. The meaning of the winding number is:
let us consider that the wave function $\psi$ is the map: $R^{3}\rightarrow
C^{1},$ when the coordinates $x$ travel around the zero point of the wave
function once, the wave function covers the zero point $W$ times. In general,
the winding number is integer number. The third term is the magnetic flux and
it is quanta. Finally, the vorticity of the total velocity is given as%
\begin{equation}
\int\left(  \mathbf{\nabla}\times\mathbf{V}\right)  \cdot d\mathbf{S}%
=\frac{e^{2}}{mc}\left[  \frac{\left(  \left\vert C_{1}\right\vert ^{2}%
\rho_{1}+\Lambda_{1}\right)  }{\rho}W_{1}\Phi_{0}+\frac{\left(  \left\vert
C_{2}\right\vert ^{2}\rho_{2}+\Lambda_{1}\right)  }{\rho}W_{2}\Phi
_{0}+W^{^{\prime\prime}}\Phi_{0}\right]  .
\end{equation}
This formula shows the velocity of the coherent state carries magnetic flux,
it is%
\begin{equation}
\Phi=\frac{\left(  \left\vert C_{1}\right\vert ^{2}\rho_{1}+\Lambda
_{1}\right)  }{\rho}W_{1}\Phi_{0}+\frac{\left(  \left\vert C_{2}\right\vert
^{2}\rho_{2}+\Lambda_{1}\right)  }{\rho}W_{2}\Phi_{0}+W^{^{\prime\prime}}%
\Phi_{0}. \label{mf}%
\end{equation}
The magnetic flux may be integer or fraction. By considering formula
(\ref{VC}), the vorticity of the total velocity should satisfy the relation
(\ref{VC}), we have%
\begin{equation}
\int\left(  \mathbf{\nabla}\times\mathbf{V}\right)  \cdot d\mathbf{S}%
=\frac{he}{m}W^{^{\prime}}.
\end{equation}
The magnetic flux from (\ref{mf}) is%
\begin{equation}
W^{^{\prime}}\Phi_{0}=\frac{\left(  \left\vert C_{1}\right\vert ^{2}\rho
_{1}+\Lambda_{1}\right)  }{\rho}W_{1}\Phi_{0}+\frac{\left(  \left\vert
C_{2}\right\vert ^{2}\rho_{2}+\Lambda_{1}\right)  }{\rho}W_{2}\Phi
_{0}+W^{^{\prime\prime}}\Phi_{0}.
\end{equation}
Then the winding number should satisfy the relation%
\begin{equation}
W^{^{\prime}}=\frac{\left(  \left\vert C_{1}\right\vert ^{2}\rho_{1}%
+\Lambda_{1}\right)  }{\rho}W_{1}+\frac{\left(  \left\vert C_{2}\right\vert
^{2}\rho_{2}+\Lambda_{1}\right)  }{\rho}W_{2}+W^{^{\prime\prime}}. \label{W}%
\end{equation}
For simple, we consider the wave function is the incoherent state, that is
$\Lambda_{1}=0$. The relation (\ref{W}) translates to%
\begin{equation}
W^{^{\prime}}-W^{^{\prime\prime}}=\frac{\left\vert C_{1}\right\vert ^{2}%
\rho_{1}}{\rho}W_{1}+\frac{\left\vert C_{2}\right\vert ^{2}\rho_{2}}{\rho
}W_{2}. \label{w1}%
\end{equation}
Let us denote $W=$ $W^{^{\prime}}-W^{^{\prime\prime}}$, (\ref{w1}) changes to%
\begin{equation}
W=\frac{\left\vert C_{1}\right\vert ^{2}\rho_{1}}{\rho}W_{1}+\frac{\left\vert
C_{2}\right\vert ^{2}\rho_{2}}{\rho}W_{2}.
\end{equation}
Based on the formula $\rho=\left\vert C_{1}\right\vert ^{2}\rho_{1}+\left\vert
C_{2}\right\vert ^{2}\rho_{2}$, so we can define%
\begin{align}
\frac{q}{p}  &  =\frac{\left\vert C_{1}\right\vert ^{2}\rho_{1}}{\rho
},\nonumber\\
\frac{p-q}{p}  &  =\frac{\left\vert C_{2}\right\vert ^{2}\rho_{2}}{\rho}.
\end{align}
where $p,q$ are integer number and $q<p$. Then the total vorticity is%
\begin{equation}
W=\frac{q}{p}W_{1}+\left(  \frac{p-q}{p}\right)  W_{2}.
\end{equation}
When the winding numbers $W_{1}=W_{2}$, we find the total winding number is
$W,$ which is integer. By considering (\ref{mf}), the magnetic flux is
integer. When $W_{1}=l_{1}p$ and $W_{2}=l_{2}p$, $l_{1},l_{2}$ are integer
number, the total winding number is $W=ql_{1}+\left(  p-q\right)  l_{2}.$ The
winding number and magnetic flux are integer. When $W_{1}$ and $W_{2}$ are
other numbers, the total winding number and magnetic flux are fractional. When
the wave function is the coherent state, let us write%
\begin{equation}
\frac{q^{^{\prime}}}{p+2q^{^{\prime}}}=\frac{\Lambda_{1}}{\rho},\text{ }%
\frac{q+q^{^{\prime}}}{p+2q^{^{\prime}}}=\frac{\left\vert C_{1}\right\vert
^{2}\rho_{1}+\Lambda_{1}}{\rho},\text{ }\frac{p-q+q^{^{\prime}}}%
{p+2q^{^{\prime}}}=\frac{\left\vert C_{2}\right\vert ^{2}\rho_{2}+\Lambda_{1}%
}{\rho}.
\end{equation}
Then the total winding number is%
\begin{equation}
W=\frac{q+q^{^{\prime}}}{p+2q^{^{\prime}}}W_{1}+\left(  \frac{p-q+q^{^{\prime
}}}{p+2q^{^{\prime}}}\right)  W_{2}.
\end{equation}
When $W_{1}=W_{2}$, the total winding number and magnetic flux are integer.
Let $W_{1}=l_{1}\left(  p+2q^{^{\prime}}\right)  $ and $W_{2}=l_{2}\left(
p+2q^{^{\prime}}\right)  ,$ the total winding number is $W=\left(
q+q^{^{\prime}}\right)  l_{1}+\left(  p-q+q^{^{\prime}}\right)  l_{2}.$ This
winding number and magnetic flux are integer. When $W_{1}$ and $W_{2}$ are
other number, the total winding number and magnetic flux are fractional.

\section{Conclusion}

In this letter, we suggest that the wave function in the two-component
superconductor is a coherent state $\Psi=C_{1}\psi_{1}+C_{2}\psi_{2}$. Where
$\psi_{1}$ and $\psi_{2}$ are the superpositions of the two condensates
respectively. Then we deduce the probability current and electronic current of
the coherent state based on Ginzburg-Landau free energy and Schr\"{o}dinger
equation. When the wave function is the coherent state, the coupling constants
are $g_{1}=\left\vert C_{1}\right\vert ^{2}+\frac{\Lambda_{1}}{\rho_{1}}$
and\ $g_{2}=\left\vert C_{2}\right\vert ^{2}+\frac{\Lambda_{1}}{\rho2}.$ At
same time, there is vector $\nabla\theta$, which can be seen as the phase of
$U(1)$ magnetic potential, arising from the coherence effects. This phase has
no contribution to the magnetic field. When the wave function is the
incoherent state, the coupling constants are $g_{1}=\left\vert C_{1}%
\right\vert ^{2}$ and\ $g_{2}=\left\vert C_{2}\right\vert ^{2}$. From these,
the ratio of the two coupling constants in the coherence is different from the
ratio in the incoherence. Then we find the reason leading to different
electronic charges is the coherence effects although there is $U(1)$ magnetic field.

The coherence effects are not only the reason leading to the different
electronic charges but also fractional vorticity and magnetic flux. In frame
of $\phi-$mapping topological theory, we find that the vorticity can be
expressed by winding number. The vorticity of the wave function can carry
magnetic flux. When the total winding number is integer, the magnetic flux is
integer. When the total winding number is fractional, the magnetic flux is fractional.\ \ \ \ \ \ \ \ \ \ \ \ \ \ \ \ \ \ \ \ \ \ \ \ \ \ \ \ \ \ \ \ \ \ \ \ \ \ \ \ \ \ \ \ \ \ \ \ \ \ \ \ \ \ \ \ \ \ \ \ \ \ \ \ \ \ \ \ \ \ \ \ \ \ \ \ \ \ \ \ \ \ \ \ \ \ \ \ \ \ \ \ \ \ \ \ \ \ \ \ \ \ \ \ \ \ \ \ \ \ \ \ \ \ \ \ \ \ \ \ \ \ \ \ \ \ \ \ \ \ \ \ \ \ \ \ \ \ \ \ \ \ \ \ \ \ \ \ \ \ \ \ \ \ \ \ \ \ \ \ \ \ \ \ \ \ \ \ \ \ \ \ \ \ \ \ \ \ \ \ \ \ \ \ \ \ \ \ \ \ \ \ \ \ \ \ \ \ \ \ \ \ \ \ \ \ \ \ \ \ \ \ \ \ \ \ \ \ \ \ \ \ \ \ \ \ \ \ \ \ \ \ \ \ \ \ \ \ \ \ \ \ \ \ \ \ \ \ \ \ \ \ \ \ \ \ \ \ \ \ \ \ \ \ \ \ \ \ \ \ \ \ \ \ \ \ \ \ \ \ \ \ \ \ \ \ \ \ \ \ \ \ \ \ \ \ \ \ \ \ \ \ \ \ \ \ \ \ \ \ \ \ \ \ \ \ \ \ \ \ \ \ \ \ \ \ \ \ \ \ \ \ \ \ \ \ \ \ \ \ \ \ \ \ \ \ \ \ \ \ \ \ \ \ \ \ \ \ \ \ \ \ \ \ \ \ \ \ \ \ \ \ \ \ \ \ \ \ \ \ \ \ \ \ \ \ \ \ \ \ \ \ \ \ \ \ \ \ \ \ \ \ \ \ \ \ \ \ \ \ \ \ \ \ \ \ \ \ \ \ \ \ \ \ \ \ \ \ \ \ \ \ \ \ \ \ \ \ \ \ \ \ \ \ \ \ \ \ \ \ \ \ \ \ \ \ \ \ \ \ \ \ \ \ \ \ \ \ \ \ \ \ \ \ \ \ \ \ \ \ \ \ \ \ \ \ \ \ \ \ \ \ \ \ \ \ \ \ \ \ \ \ \ \ \ \ \ \ \ \ \ \ \ \ \ \ \ \ \ \ \ \ \ \ \ \ \ \ \ \ \ \ \ \ \ \ \ \ \ \ \ \ \ \ \ \ \ \ \ \ \ \ \ \ \ \ \ \ \ \ \ \ \ \ \ \ \ \ \ \ \ \ \ \ \ \ \ \ \ \ \ \ \ \ \ \ \ \ \ \ \ \ \ \ \ \ \ \ \ \ \ \ \ \ \ \ \ \ \ \ \ \ \ \ \ \ \ \ \ \ \ \ \ \ \ \ \ \ \ \ \ \ \ \ \ \ \ \ \ \ \ \ \ \ \ \ \ \ \ \ \ \ \ \ \ \ \ \ \ \ \ \ \ \ \ \ \ \ \ \ \ \ \ \ \ \ \ \ \ \ \ \ \ \ \ \ \ \ \ \ \ \ \ \ \ \ \ \ \ \ \ \ \ \ \ \ \ \ \ \ \ \ \ \ \ \ \ \ \ \ \ \ \ \ \ \ \ \ \ \ \ \ \ \ \ \ \ \ \ \ \ \ \ \ \ \ \ \ \ \ \ \ \ \ \ \ \ \ \ \ \ \ \ \ \ \ \ \ \ \ \ \ \ \ \ \ \ \ \ \ \ \ \ \ \ \ \ \ \ \ \ \ \ \ \ \ \ \ \ \ \ \ \ \ \ \ \ \ \ \ \ \ \ \ \ \ \ \ \ \ \ \ \ \ \ \ \ \ \ \ \ \ \ \ \ \ \ \ \ \ \ \ \ \ \ \ \ \ \ \ \ \ \ \ \ \ \ \ \ \ \ \ \ \ \ \ \ \ \ \ \ \ \ \ \ \ \ \ \ \ \ \ \ \ \ \ \ \ \ \ \ \ \ \ \ \ \ \ \ \ \ \ \ \ \ \ \ \ \ \ \ \ \ \ \ \ \ \ \ \ \ \ \ \ \ \ \ \ \ \ \ \ \ \ \ \ \ \ \ \ \ \ \ \ \ \ \ \ \ \ \ \ \ \ \ \ \ \ \ \ \ \ \ \ \ \ \ \ \ \ \ \ \ \ \ \ \ \ \ \ \ \ \ \ \ \ \ \ \ \ \ \ \ \ \ \ \ \ \ \ \ \ \ \ \ \ \ \ \ \ \ \ \ \ \ \ \ \ \ \ \ \ \ \ \ \ \ \ \ \ \ \ \ \ \ \ \ \ \ \ \ \ \ \ \ \ \ \ \ \ \ \ \ \ \ \ \ \ \ \ \ \ \ \ \ \ \ \ \ \ \ \ \ \ \ \ \ \ \ \ \ \ \ \ \ \ \ \ \ \ \ \ \ \ \ \ \ \ \ \ \ \ \ \ \ \ \ \ \ \ \ \ \ \ \ \ \ \ \ \ \ \ \ \ \ \ \ \ \ \ \ \ \ \ \ \ \ \ \ \ \ \ \ \ \ \ \ \ \ \ \ \ \ \ \ \ \ \ \ \ \ \ \ \ \ \ \ \ \ \ \ \ \ \ \ \ \ \ \ \ \ \ \ \ \ \ \ \ \ \ \ \ \ \ \ \ \ \ \ \ \ \ \ \ \ \ \ \ \ \ \ \ \ \ \ \ \ \ \ \ \ \ \ \ \ \ \ \ \ \ \ \ \ \ \ \ \ \ \ \ \ \ \ \ \ \ \ \ \ \ \ \ \ \ \ \ \ \ \ \ \ \ \ \ \ \ \ \ \ \ \ \ \ \ \ \ \ \ \ \ \ \ \ \ \ \ \ \ \ \ \ \ \ \ \ \ \ \ \ \ \ \ \ \ \ \ \ \ \ \ \ \ \ \ \ \ \ \ \ \ \ \ \ \ \ \ \ \ \ \ \ \ \ \ \ \ \ \ \ \ \ \ \ \ \ \ \ \ \ \ \ \ \ \ \ \ \ \ \ \ \ \ \ \ \ \ \ \ \ \ \ \ \ \ \ \ \ \ \ \ \ \ \ \ \ \ \ \ \ \ \ \ \ \ \ \ \ \ \ \ \ \ \ \ \ \ \ \ \ \ \ \ \ \ \ \ \ \ \ \ \ \ \ \ \ \ \ \ \ \ \ \ \ \ \ \ \ \ \ \ \ \ \ \ \ \ \ \ \ \ \ \ \ \ \ \ \ \ \ \ \ \ \ \ \ \ \ \ \ \ \ \ \ \ \ \ \ \ \ \ \ \ \ \ \ \ \ \ \ \ \ \ \ \ \ \ \ \ \ \ \ \ \ \ \ \ \ \ \ \ \ \ \ \ \ \ \ \ \ \ \ \ \ \ \ \ \ \ \ \ \ \ \ \ \ \ \ \ \ \ \ \ \ \ \ \ \ \ \ \ \ \ \ \ \ \ \ \ \ \ \ \ \ \ \ \ \ \ \ \ \ \ \ \ \ \ \ \ \ \ \ \ \ \ \ \ \ \ \ \ \ \ \ \ \ \ \ \ \ \ \ \ \ \ \ \ \ \ \ \ \ \ \ \ \ \ \ \ \ \ \ \ \ \ \ \ \ \ \ \ \ \ \ \ \ \ \ \ \ \ \ \ \ \ \ \ \ \ \ \ \ \ \ \ \ \ \ \ \ \ \ \ \ \ \ \ \ \ \ \ \ \ \ \ \ \ \ \ \ \ \ \ \ \ \ \ \ \ \ \ \ \ \ \ \ \ \ \ \ \ \ \ \ \ \ \ \ \ \ \ \ \ \ \ \ \ \ \ \ \ \ \ \ \ \ \ \ \ \ \ \ \ \ \ \ \ \ \ \ \ \ \ \ \ \ \ \ \ \ \ \ \ \ \ \ \ \ \ \ \ \ \ \ \ \ \ \ \ \ \ \ \ \ \ \ \ \ \ \ \ \ \ \ \ \ \ \ \ \ \ \ \ \ \ \ \ \ \ \ \ \ \ \ \ \ \ \ \ \ \ \ \ \ \ \ \ \ \ \ \ \ \ \ \ \ \ \ \ \ \ \ \ \ \ \ \ \ \ \ \ \ \ \ \ \ \ \ \ \ \ \ \ \ \ \ \ \ \ \ \ \ \ \ \ \ \ \ \ \ \ \ \ \ \ \ \ \ \ \ \ \ \ \ \ \ \ \ \ \ \ \ \ \ \ \ \ \ \ \ \ \ \ \ \ \ \ \ \ \ \ \ \ \ \ \ \ \ \ \ \ \ \ \ \ \ \ \ \ \ \ \ \ \ \ \ \ \ \ \ \ \ \ \ \ \ \ \ \ \ \ \ \ \ \ \ \ \ \ \ \ \ \ \ \ \ \ \ \ \ \ \ \ \ \ \ \ \ \ \ \ \ \ \ \ \ \ \ \ \ \ \ \ \ \ \ \ \ \ \ \ \ \ \ \ \ \ \ \ \ \ \ \ \ \ \ \ \ \ \ \ \ \ \ \ \ \ \ \ \ \ \ \ \ \ \ \ \ \ \ \ \ \ \ \ \ \ \ \ \ \ \ \ \ \ \ \ \ \ \ \ \ \ \ \ \ \ \ \ \ \ \ \ \ \ \ \ \ \ \ \ \ \ \ \ \ \ \ \ \ \ \ \ \ \ \ \ \ \ \ \ \ \ \ \ \ \ \ \ \ \ \ \ \ \ \ \ \ \ \ \ \ \ \ \ \ \ \ \ \ \ \ \ \ \ \ \ \ \ \ \ \ \ \ \ \ \ \ \ \ \ \ \ \ \ \ \ \ \ \ \ \ \ \ \ \ \ \ \ \ \ \ \ \ \ \ \ \ \ \ \ \ \ \ \ \ \ \ \ \ \ \ \ \ \ \ \ \ \ \ \ \ \ \ \ \ \ \ \ \ \ \ \ \ \ \ \ \ \ \ \ \ \ \ \ \ \ \ \ \ \ \ \ \ \ \ \ \ \ \ \ \ \ \ \ \ \ \ \ \ \ \ \ \ \ \ \ \ \ \ \ \ \ \ \ \ \ \ \ \ \ \ \ \ \ \ \ \ \ \ \ \ \ \ \ \ \ \ \ \ \ \ \ \ \ \ \ \ \ \ \ \ \ \ \ \ \ \ \ \ \ \ \ \ \ \ \ \ \ \ \ \ \ \ \ \ \ \ \ \ \ \ \ \ \ \ \ \ \ \ \ \ \ \ \ \ \ \ \ \ \ \ \ \ \ \ \ \ \ \ \ \ \ \ \ \ \ \ \ \ \ \ \ \ \ \ \ \ \ \ \ \ \ \ \ \ \ \ \ \ \ \ \ \ \ \ \ \ \ \ \ \ \ \ \ \ \ \ \ \ \ \ \ \ \ \ \ \ \ \ \ \ \ \ \ \ \ \ \ \ \ \ \ \ \ \ \ \ \ \ \ \ \ \ \ \ \ \ \ \ \ \ \ \ \ \ \ \ \ \ \ \ \ \ \ \ \ \ \ \ \ \ \ \ \ \ \ \ \ \ \ \ \ \ \ \ \ \ \ \ \ \ \ \ \ \ \ \ \ \ \ \ \ \ \ \ \ \ \ \ \ \ \ \ \ \ \ \ \ \ \ \ \ \ \ \ \ \ \ \ \ \ \ \ \ \ \ \ \ \ \ \ \ \ \ \ \ \ \ \ \ \ \ \ \ \ \ \ \ \ \ \ \ \ \ \ \ \ \ \ \ \ \ \ \ \ \ \ \ \ \ \ \ \ \ \ \ \ \ \ \ \ \ \ \ \ \ \ \ \ \ \ \ \ \ \ \ \ \ \ \ \ \ \ \ \ \ \ \ \ \ \ \ \ \ \ \ \ \ \ \ \ \ \ \ \ \ \ \ \ \ \ \ \ \ \ \ \ \ \ \ \ \ \ \ \ \ \ \ \ \ \ \ \ \ \ \ \ \ \ \ \ \ \ \ \ \ \ \ \ \ \ \ \ \ \ \ \ \ \ \ \ \ \ \ \ \ \ \ \ \ \ \ \ \ \ \ \ \ \ \ \ \ \ \ \ \ \ \ \ \ \ \ \ \ \ \ \ \ \ \ \ \ \ \ \ \ \ \ \ \ \ \ \ \ \ \ \ \ \ \ \ \ \ \ \ \ \ \ \ \ \ \ \ \ \ \ \ \ \ \ \ \ \ \ \ \ \ \ \ \ \ \ \ \ \ \ \ \ \ \ \ \ \ \ \ \ \ \ \ \ \ \ \ \ \ \ \ \ \ \ \ \ \ \ \ \ \ \ \ \ \ \ \ \ \ \ \ \ \ \ \ \ \ \ \ \ \ \ \ \ \ \ \ \ \ \ \ \ \ \ \ \ \ \ \ \ \ \ \ \ \ \ \ \ \ \ \ \ \ \ \ \ \ \ \ \ \ \ \ \ \ \ \ \ \ \ \ \ \ \ \ \ \ \ \ \ \ \ \ \ \ \ \ \ \ \ \ \ \ \ \ \ \ \ \ \ \ \ \ \ \ \ \ \ \ \ \ \ \ \ \ \ \ \ \ \ \ \ \ \ \ \ \ \ \ \ \ \ \ \ \ \ \ \ \ \ \ \ \ \ \ \ \ \ \ \ \ \ \ \ \ \ \ \ \ \ \ \ \ \ \ \ \ \ \ \ \ \ \ \ \ \ \ \ \ \ \ \ \ \ \ \ \ \ \ \ \ \ \ \ \ \ \ \ \ \ \ \ \ \ \ \ \ \ \ \ \ \ \ \ \ \ \ \ \ \ \ \ \ \ \ \ \ \ \ \ \ \ \ \ \ \ \ \ \ \ \ \ \ \ \ \ \ \ \ \ \ \ \ \ \ \ \ \ \ \ \ \ \ \ \ \ \ \ \ \ \ \ \ \ \ \ \ \ \ \ \ \ \ \ \ \ \ \ \ \ \ \ \ \ \ \ \ \ \ \ \ \ \ \ \ \ \ \ \ \ \ \ \ \ \ \ \ \ \ \ \ \ \ \ \ \ \ \ \ \ \ \ \ \ \ \ \ \ \ \ \ \ \ \ \ \ \ \ \ \ \ \ \ \ \ \ \ \ \ \ \ \ \ \ \ \ \ \ \ \ \ \ \ \ \ \ \ \ \ \ \ \ \ \ \ \ \ \ \ \ \ \ \ \ \ \ \ \ \ \ \ \ \ \ \ \ \ \ \ \ \ \ \ \ \ \ \ \ \ \ \ \ \ \ \ \ \ \ \ \ \ \ \ \ \ \ \ \ \ \ \ \ \ \ \ \ \ \ \ \ \ \ \ \ \ \ \ \ \ \ \ \ \ \ \ \ \ \ \ \ \ \ \ \ \ \ \ \ \ \ \ \ \ \ \ \ \ \ \ \ \ \ \ \ \ \ \ \ \ \ \ \ \ \ \ \ \ \ \ \ \ \ \ \ \ \ \ \ \ \ \ \ \ \ \ \ \ \ \ \ \ \ \ \ \ \ \ \ \ \ \ \ \ \ \ \ \ \ \ \ \ \ \ \ \ \ \ \ \ \ \ \ \ \ \ \ \ \ \ \ \ \ \ \ \ \ \ \ \ \ \ \ \ \ \ \ \ \ \ \ \ \ \ \ \ \ \ \ \ \ \ \ \ \ \ \ \ \ \ \ \ \ \ \ \ \ \ \ \ \ \ \ \ \ \ \ \ \ \ \ \ \ \ \ \ \ \ \ \ \ \ \ \ \ \ \ \ \ \ \ \ \ \ \ \ \ \ \ \ \ \ \ \ \ \ \ \ \ \ \ \ \ \ \ \ \ \ \ \ \ \ \ \ \ \ \ \ \ \ \ \ \ \ \ \ \ \ \ \ \ \ \ \ \ \ \ \ \ \ \ \ \ \ \ \ \ \ \ \ \ \ \ \ \ \ \ \ \ \ \ \ \ \ \ \ \ \ \ \ \ \ \ \ \ \ \ \ \ \ \ \ \ \ \ \ \ \ \ \ \ \ \ \ \ \ \ \ \ \ \ \ \ \ \ \ \ \ \ \ \ \ \ \ \ \ \ \ \ \ \ \ \ \ \ \ \ \ \ \ \ \ \ \ \ \ \ \ \ \ \ \ \ \ \ \ \ \ \ \ \ \ \ \ \ \ \ \ \ \ \ \ \ \ \ \ \ \ \ \ \ \ \ \ \ \ \ \ \ \ \ \ \ \ \ \ \ \ \ \ \ \ \ \ \ \ \ \ \ \ \ \ \ \ \ \ \ \ \ \ \ \ \ \ \ \ \ \ \ \ \ \ \ \ \ \ \ \ \ \ \ \ \ \ \ \ \ \ \ \ \ \ \ \ \ \ \ \ \ \ \ \ \ \ \ \ \ \ \ \ \ \ \ \ \ \ \ \ \ \ \ \ \ \ \ \ \ \ \ \ \ \ \ \ \ \ \ \ \ \ \ \ \ \ \ \ \ \ \ \ \ \ \ \ \ \ \ \ \ \ \ \ \ \ \ \ \ \ \ \ \ \ \ \ \ \ \ \ \ \ \ \ \ \ \ \ \ \ \ \ \ \ \ \ \ \ \ \ \ \ \ \ \ \ \ \ \ \ \ \ \ \ \ \ \ \ \ \ \ \ \ \ \ \ \ \ \ \ \ \ \ \ \ \ \ \ \ \ \ \ \ \ \ \ \ \ \ \ \ \ \ \ \ \ \ \ \ \ \ \ \ \ \ \ \ \ \ \ \ \ \ \ \ \ \ \ \ \ \ \ \ \ \ \ \ \ \ \ \ \ \ \ \ \ \ \ \ \ \ \ \ \ \ \ \ \ \ \ \ \ \ \ \ \ \ \ \ \ \ \ \ \ \ \ \ \ \ \ \ \ \ \ \ \ \ \ \ \ \ \ \ \ \ \ \ \ \ \ \ \ \ \ \ \ \ \ \ \ \ \ \ \ \ \ \ \ \ \ \ \ \ \ \ \ \ \ \ \ \ \ \ \ \ \ \ \ \ \ \ \ \ \ \ \ \ \ \ \ \ \ \ \ \ \ \ \ \ \ \ \ \ \ \ \ \ \ \ \ \ \ \ \ \ \ \ \ \ \ \ \ \ \ \ \ \ \ \ \ \ \ \ \ \ \ \ \ \ \ \ \ \ \ \ \ \ \ \ \ \ \ \ \ \ \ \ \ \ \ \ \ \ \ \ \ \ \ \ \ \ \ \ \ \ \ \ \ \ \ \ \ \ \ \ \ \ \ \ \ \ \ \ \ \ \ \ \ \ \ \ \ \ \ \ \ \ \ \ \ \ \ \ \ \ \ \ \ \ \ \ \ \ \ \ \ \ \ \ \ \ \ \ \ \ \ \ \ \ \ \ \ \ \ \ \ \ \ \ \ \ \ \ \ \ \ \ \ \ \ \ \ \ \ \ \ \ \ \ \ \ \ \ \ \ \ \ \ \ \ \ \ \ \ \ \ \ \ \ \ \ \ \ \ \ \ \ \ \ \ \ \ \ \ \ \ \ \ \ \ \ \ \ \ \ \ \ \ \ \ \ \ \ \ \ \ \ \ \ \ \ \ \ \ \ \ \ \ \ \ \ \ \ \ \ \ \ \ \ \ \ \ \ \ \ \ \ \ \ \ \ \ \ \ \ \ \ \ \ \ \ \ \ \ \ \ \ \ \ \ \ \ \ \ \ \ \ \ \ \ \ \ \ \ \ \ \ \ \ \ \ \ \ \ \ \ \ \ \ \ \ \ \ \ \ \ \ \ \ \ \ \ \ \ \ \ \ \ \ \ \ \ \ \ \ \ \ \ \ \ \ \ \ \ \ \ \ \ \ \ \ \ \ \ \ \ \ \ \ \ \ \ \ \ \ \ \ \ \ \ \ \ \ \ \ \ \ \ \ \ \ \ \ \ \ \ \ \ \ \ \ \ \ \ \ \ \ \ \ \ \ \ \ \ \ \ \ \ \ \ \ \ \ \ \ \ \ \ \ \ \ \ \ \ \ \ \ \ \ \ \ \ \ \ \ \ \ \ \ \ \ \ \ \ \ \ \ \ \ \ \ \ \ \ \ \ \ \ \ \ \ \ \ \ \ \ \ \ \ \ \ \ \ \ \ \ \ \ \ \ \ \ \ \ \ \ \ \ \ \ \ \ \ \ \ \ \ \ \ \ \ \ \ \ \ \ \ \ \ \ \ \ \ \ \ \ \ \ \ \ \ \ \ \ \ \ \ \ \ \ \ \ \ \ \ \ \ \ \ \ \ \ \ \ \ \ \ \ \ \ \ \ \ \ \ \ \ \ \ \ \ \ 

\begin{acknowledgement}
Supported by \textquotedblleft the Fundamental Research Funds for the Central
Universities NO.YX2013-21\newline
\end{acknowledgement}


\begin{thebibliography}{99}                                                                                               %


\bibitem {two component}J.Garaud, K.A.Sellin, J.J\"{a}ykk\"{a}, E.Babaev,
Phys. Rev. B 89 (2014) 104508

\bibitem {two component1}M.Cipriani, M.Nitta, Phys. Rev. Lett. 111 (2013) 170401

\bibitem {two component2}K.Kasamatsu, H.Takeuchi, M.Tsubota, M.Nitta, Phys.
Rev. A 88 (2013) 013620

\bibitem {twocom}A.M.Kamchatnov, Y.V.Kartashov, Phys. Rev. Lett. 111 (2013) 140402

\bibitem {MgB}J.Nagamatsu, N.Nakagawa, T.Muranaka, Y.Zenitani, J.Akimitsu,
Nature, 410 (2001) 63

\bibitem {MgB1}F.Bouquet, R.A.Fisher, N.E.Phillips, D.G.Hinks, J.D.Jorgensen,
Phys. Rev. Lett. 87 (2001) 047001

\bibitem {MgB2}P.Szab\'{o}, P.Samuely, J.Ka\v{c}mar\v{c}\'{\i}k, T.Klein,
J.Marcus, D.Fruchart, S.Miraglia, C.Marcenat, A.G.M.Jansen, Phys. Rev. Lett.
87 (2001) 137005

\bibitem {compositV}E.Babaev, L.D.Faddeev, A.Niemi, Phys. Rev. B 65 (2002) 100512

\bibitem {compositV1}E.Babaev, Phys. Rev. Lett. 89 (2002) 067001

\bibitem {compositV2}Y.S.Duan, X.G.Shi, Int. J. Mod. Phys. B 18 (2002) 1309

\bibitem {Typ1.5}M.Silaev, E.Babaev, Phys. Rev. B 84 (2011) 094515

\bibitem {Typ1.51}E.Babaev, J.Carlastr\"{o}m, JGaraud, M.Silaev, J.M.Speight,
Physica C 479 (2012) 2

\bibitem {different charges}J.Garaud, E.Babaev, Phys. Rev. B 89 (2014) 214507

\bibitem {JE}X.Shi, Phys. Scr. 85 (2012) 055005

\bibitem {topo}Y.S.Duan, M.L.Ge, Sci Sin. 11 (1979) 1072

\bibitem {topo1}X.Shi, Y.S.Duan, Mod. Phys. Lett. A 26 (2011) 1363

\bibitem {topo2}X.Shi, M.Yu, Y.S.Duan, Mod. Phys. Lett. A 24 (2009) 453

\bibitem {duan}L.B.Fu, Y.S.Duan, H.Zhang, Phys. Rev. D 61 (2000) 045004
\end{thebibliography}
\end{document}